\documentclass[12pt,headings=small]{scrartcl}
\usepackage[authoryear]{natbib}
\usepackage{helvet}
\usepackage{mathptmx}

\usepackage[british]{babel}
\usepackage{graphicx}
\usepackage{color}

\newif\ifreview
\reviewtrue
\reviewfalse

\ifreview
\input{reviewpreamble}
\fi

\usepackage[colorlinks%
	,bookmarks,bookmarksnumbered%
	,breaklinks,hyperindex%
	,linkcolor={blue}%
	,citecolor={blue}%
	,anchorcolor={blue}%
	,pagecolor={blue}%
	,menucolor={blue}%
	,urlcolor={blue}%
	,pdfauthor={Carsten Lemmen <carsten.lemmen@hzg.de>}%
	,pdfsubject={Malthusian assumptions, Boserupian response in models of the transitions to agriculture}%
	,pdfkeywords={Early agriculture, Ester Boserup, adaptation},
	,pdftitle={Lemmen2012_boserup_tex.tex}
	]{hyperref}%

\newcommand{\executeiffilenewer}[3]{%
\ifnum\pdfstrcmp{\pdffilemoddate{#1}}%
{\pdffilemoddate{#2}}>0%
{\immediate\write18{#3}}\fi%
}

\newcommand{\refeq}[1]{equation~\ref{#1}}
\newcommand{\refeqs}[2]{equations~\ref{#1}--\ref{#2}}

\newcommand{\reffig}[1]{Figure~\ref{#1}}

\newcommand{\comment}[1]{}

\newcommand{\dd}[2]{\ensuremath{\frac{\mathrm{d}#1}{\mathrm{d}#2}}}
\newcommand{\del}[2]{\ensuremath{\frac{\partial{}#1}{\partial{}#2}}}
\newcommand{\si}{\ensuremath{\mathrm{SI}}}

\title{
\ifreview\normalfont\MakeUppercase\fi\Large\bfseries Malthusian assumptions, Boserupian response in models of the transitions to agriculture
\ifreview\vfill\else
\footnote{To appear in: ``Society, Nature and History: The Legacy of Ester Boserup'', Springer, Vienna}
\fi
}

\ifreview\subtitle{\normalfont\rmfamily by Carsten Lemmen\vfill}
\author{\parbox{\hsize}{\centering\large\bfseries\hrulefill\\ 
DO NOT CITE IN ANY CONTEXT WITHOUT PERMISSION
\\ \hrulefill\vfill
}}
\else\author{Carsten Lemmen}
\fi

\date{\ifreview\vfill\large\raggedright\else\small\fi Institut f\"ur K\"ustenforschung, Helmholtz-Zentrum Geesthacht, Max-Planck Stra\ss e~1, 21501~Geesthacht, Germany (\href{mailto:carsten.lemmen@hzg.de}{carsten.lemmen@hzg.de})
\ifreview\vfill\small In: Society, Nature and History: The Legacy of Ester Boserup, Springer, Vienna\fi
}

\begin{document}
\maketitle

\ifreview\newpage\fi\paragraph*{Abstract.}

In the many transitions from foraging to agropastoralism it is debated whether the primary drivers are innovations in technology or increases of population. The driver discussion traditionally separates Malthusian (technology driven) from Boserupian (population driven) theories. I present a numerical model of the transitions to agriculture and discuss this model in the light of the population versus technology debate and in Boserup's analytical framework in development theory. Although my model is based on ecological---Neomalthusian---principles, the coevolutionary positive feedback relationship between technology and population results in a seemingly Boserupian response: innovation is greatest when population pressure is highest. This outcome is not only visible in the theory-driven reduced model, but is also present in a corresponding ``real world'' simulator which was tested against archaeological data, demonstrating the relevance and validity of the coevolutionary model. The lesson to be learned is that not all that acts Boserupian needs Boserup at its core.

\ifreview\newpage\else\fi
\section{Transitions to agriculture}

% Global impact of transition
The relationship between humans and their environment underwent a radical change during the last 10,000~years: from mobile and small groups of foraging people to sedentary extensive cultivators and on to high-density intensive agriculture modern society; these transitions fundamentally turned the formerly predominantly passive human user of the environment into an active component of the Earth system. The most striking global impact is only visible and measurable during the last 150~years \citep{Crutzen2000a,Crutzen2002gm}; much earlier, however, the use of forest resources for metal smelting from early Roman times and the medieval extensive agricultural system had already changed the landscape \citep{Barker2011,Kaplan2009}; global climate effects of these early extensive cultivation and harvesting practices are yet under debate \citep{Ruddiman2003,Lemmen2010,Kaplan2011,Stocker2011}.

% Heterogeneity of transition
% Transition to agriculture in stages
Transitions to agriculture occurred in almost every region of the world, earliest in China and the Near East over 9000 years ago  \citep[][]{Kuijt2002,Londo2006}, and latest in Australia and Oceania with the arrival of Polynesian and European immigrants few hundred years ago \citep{Diamond2003flf}. While each local transition can be considered revolutionary, the many diverse mechanisms, environments, and cultural contexts of each agricultural transition make it difficult to speak of the one 'Neolithic revolution', as the transition to farming and herding was termed by V.\,G.\,Childe almost a century ago \citep{Childe1925}. The transitions from foraging to farming were not only one big step, but may have consisted of intermediary stages: \citet{Bogaard2005gan} looks at the transition in terms of the land use system: she sees first inadvertent cultivation then horticulture then simple and then advanced agriculture, while \citet{Boserup1965} discriminates these stages by the management practice ranging from forest, bush and short fallow to annual and multi cropping. 

% Innovation, intensification, population
In contemporary hunting-gathering societies much less time has to be devoted to procuring food from hunting and gathering opposed to agriculture and herding  \citep[e.g.,][]{Sahlins1972}; less labor is required for long fallow systems compared to intensive multi-cropping agriculture\citep{Boserup1965}. So why farm? Different responses from archaeology \citep{Barker2011}, demography \citep{Turchin2009}, historical economy \citep{Weisdorf2005fff}, and ecosystem modeling \citep{Wirtz2003gdm} call upon processes such as social reorganization, the value of leisure, changing resources, or coevolutionary thresholds. 

The probably simplest relationship was proposed by  \citet[][p.11]{Malthus1798}, namely that more production sustains larger population. With larger population, more production is possible, thereby constituting a positive feedback loop, which ideally results in ever greater (geometric) growth and productivity. That this is not the case in a world with finite resources was expressed by \citet[][p.~4]{Malthus1798} by stating that ``Population, when unchecked, increases at a geometrical ratio. Subsistence increases only in an arithmetical ratio. A slight acquaintance with numbers will show the immensity of the first power in comparison with the second''. Malthus identified the need for positive and preventive checks to balance population increase with the limited capacity of resources.

% Extensive and intensive Increase in subsistence
How does an increase in productivity come about? First and foremost, the input of more labor increases productivity \citep[][p.~11]{Malthus1798}, subject to the constraints of finite resources and diminishing returns. Where Malthus, however, focused on extensive productivity increase, the intensification component of productivity increase was highlighted by  \citet{Boserup1965}. Investments in a more intensive production system would, however, require large additional labor, and the benefits of such investments were often small. To stimulate an investment in more intensive agriculture, Boserup requires population pressure.

% Innovation
Both \citet{Malthus1798,Malthus1826}  and \citet{Boserup1965,Boserup1981} concentrate on the role of labor (and later division of labor and social/family organization) and innovations which increase area productivity (like storage or tools, requiring relatively more labor for harvesting, building, and tool processing). Both authors neglect the role of labor-independent innovation, or innovations which increase both area and labor productivity; these are innovations in the resources themselves, such as cultivation of higher-yielding grains or imported high yield varieties, or their management such as water rights; this distinction may not be unambiguous for all innovations, it is used here conceptually. Labor-independent innovation can be stimulated by diversity and density of a population, both of which are positively related to population size. Already \citet[][p.~156]{Darwin1859} wrote ``The more diversified [..], by so much will they be better enabled to seize on many and widely diversified places in the polity of nature''. Translated into the realm of innovativity, Darwin's ``seizing of places'', or niche occupation, would be the realization of technical and scientific opportunities. As for density as a stimulus of innovation, it is aggregation which constitutes a motor of technological and cultural change \citep{Smith1776,Boyd1995}\footnote{This does not, however, give the reason for a particular choice of one innovation over another \citep{Sober1992}.}.

\section{Models of population, production, and innovation}

In 1996, E.\,Boserup reflected on the problems arising from the differences in terminology and methodology when comparing different models of development theories \citep{Boserup1996}. She suggested a common framework to facilitate interdisciplinary cooperation based on six structures: Environment ($E$), Population ($P$), technology, occupational structure, family structure and culture. In this framework, she then interpreted the major works of Adam Smith, Thomas Malthus, Max Weber, Karl Marx, David Ricardo, and Neomalthusian thinking, as well as her own view on different stages of the developmental process.

\begin{figure}
\centering
\includegraphics[width=\ifreview0.8\else0.8\fi\hsize]{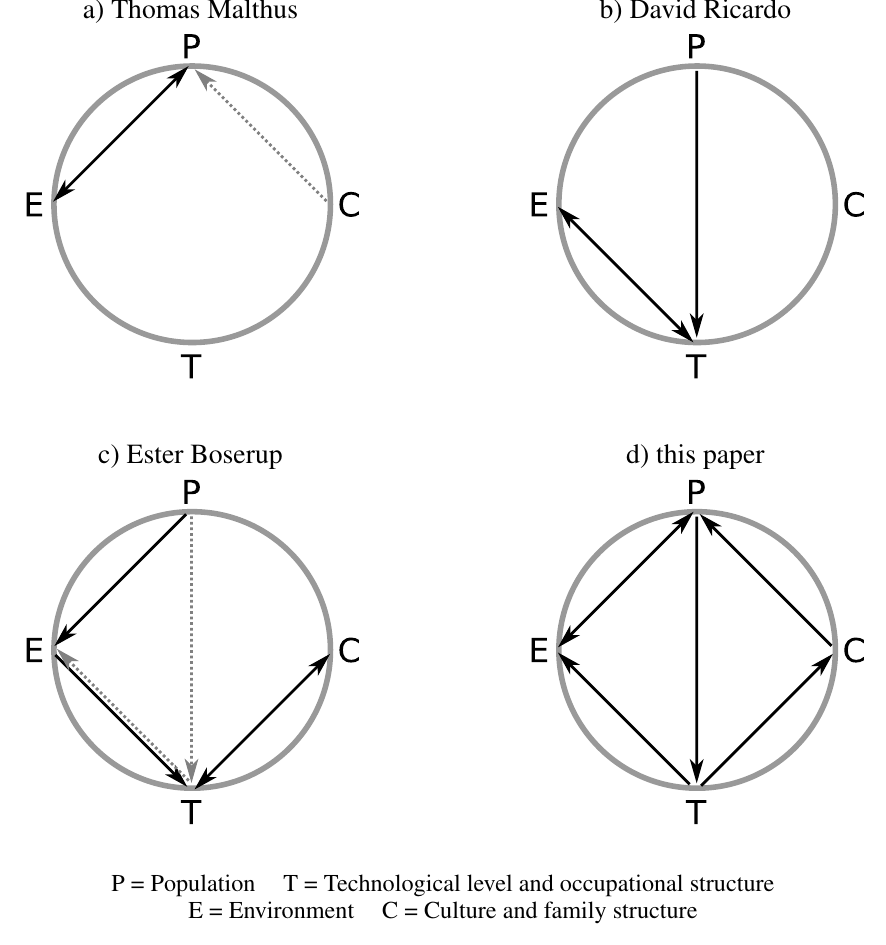}
\caption{Four compartment framework for the interrelationship between population, environment, technology, and culture.  Four economic theories are contrasted: the essays on the principles of population by T.~\citet{Malthus1798,Malthus1826} (panel a, dotted line indicates the revised essay including culture change);   D.~\citet{Ricardo1821}'s principles of economy (panel b); E.~\citet{Boserup1965,Boserup1981}'s theories for five transitions explained in \citet{Boserup1996} (panel c, \citeyear{Boserup1981} refinements shown as dotted lines); and the ecological model proposed in this chapter.   The framework is a simplification of the six compartment framework originally proposed by \citet{Boserup1996}.}
\label{fig:models}
\end{figure}

For many of the theories and models discussed by Boserup in this framework, the partitioning in six structures can be simplified by (a) aggregating technology and occupational structure into a single entity technology ($T$), and by (b) aggregating culture and family structure into a single entity culture ($C$). Aggregating technology and occupational structure means that I assume here that changes in technology are equivalent to changes in organization and that the location of technological change is the occupational sector. By aggregating family structure and culture I assume that values and social conventions penetrate from the society into the family and are governed by similar dynamics. The reduced framework then consists of the four compartments population, environment, technology, and culture (\textit{PETC},  \reffig{fig:models}).

% Discuss why this is good.
% Reviewer says You can only aggregate culture and familiy, if most of the agricultural production happens within family structures, I don't agree
In this \textit{PETC} framework, the one referring to \citet{Malthus1798} involves only population and environment. Population growth exerts pressure on the environment, and failure to provide adequate resources from the environment acts as a positive check on population through higher mortality (\reffig{fig:models}a). Technology does not play a role in this simplest Malthusian model\footnote{Malthus considered the increase of carrying capacity by autonomously occurring inventions \citep{Lee1986}, however, this was not discussed by \citet{Boserup1996} in her model intercomparison.}. Culture in the form of preventive checks---such as birth control---acts on population only in later versions of his theory  \citep{Malthus1826}. At its core remains ``the dependent role he assigns to population growth'' \citep{Marquette1997}. D.\,\citet{Ricardo1821} proposed that the incentive to intensify and develop technologies comes from a stimulus in population pressure. The demand for more land ($E$), however, leads to declining marginal benefits of and a negative feedback on innovation ($T$) due to high costs of renting the land (\reffig{fig:models}b). In Ricardo's work, population is independent, and technology and environment are the dependent variables.

Population is also the driving factor in Boserup's (\citeyear{Boserup1965,Boserup1981}) works. Of the six transitions considered by \citet{Boserup1996}, five can be accommodated within my \textit{PETC} framework as a succession of population, environment, technology, and culture: foraging to crop production, village development, Eastern hemisphere pastoralism, urbanization, and industrialization (\reffig{fig:models}c)\footnote{The sixth transition---Western European fertility decline---follows a different path as a succession of technology, environment, culture, and last population; it is not considered here.}. In all these transitions, population growth leads to pressure felt from the limited environmental resources, which in turn stimulates technological and organizational change, and later results in cultural changes evident in cults, social hierarchies, women's status, and status symbols. Within this group of five transitions, her model of village development, in addition, has a direct population--technology link, and allows for a feedback of the land resources on occupational structure (dotted lines in \reffig{fig:models}c). Furthermore, her model of the foraging to farming transition includes a feedback from culture to organizational structure (not shown).

\section{A combined model and `real' world application}

I suggest here a different model of population development taking the foraging to farming transition as an example (\reffig{fig:models}d). This model is a reduced form of the Global Land Use and technological Evolution Simulator (GLUES, described below), which has been operationally applied to a number of problems in archaeology and climate research \citep{Kaplan2011,Lemmen2010,Lemmen2012,Lemmen2011}. The reduced model shares the functional characteristics of the full model, but it is not spatially explicit and the biogeographic and climate background is regarded as constant (see Appendix for equations).

In terms of the \textit{PETC} framework, the dynamics between population, environment, technology and culture is the following (\reffig{fig:models}d, cmp.~\citealt{Boserup1996}, p.\,509).

 \begin{enumerate}
 \item P$\rightarrow$ T$\rightarrow$ P\quad Population growth stimulates innovation by aggregation and diversity.  Innovations in, e.g., health care increase population;
 \item P$\rightarrow$ E $\rightarrow$ P\quad  Population growth uses ever more land for hunting and exerts pressure on the game stock, higher population densities damage the environment, and food shortage leads to reduced fertility (preventive check) or higher mortality (positive check).  The rising capacity of the environment supports higher population;
 \item T$\rightarrow$ E\quad  More intensive foraging or farming strategies damage the environment, while efficiency gains lead to higher capacity of the environment;
 \item T$\rightarrow$ C\quad  Adoption of novel technologies induces changes in social structure where specialists and leaders or cults emerge;
\item C$\rightarrow$ P\quad  Family and social structure change reproduction rates.
\end{enumerate}
 
\citet{Richerson1998} claim that basically all models which are rooted in ecology are Neomalthusian in essence, i.e., they can be characterized by a P$\rightarrow$\ T\ $\rightarrow$\ E loop in \citeauthor{Boserup1996}'s (\citeyear{Boserup1996}) framework. This loop can be detected in my model, as well; in fact, historically it developed from ecosystem models of tree stands or algal communities\citep{Wirtz1996eve}. Unlike many other models, however, GLUES is based on coevolutionary dynamics of technologies and population, and as such has no a priori information on whether there is a (Malthusian) ``invention-pull view of population history'' \citep[][p. 98]{Lee1986}, or whether population is the (Boserupian) driver of development\footnote{See also  \citet{Simon1993} for a detailed discussion.}. Applications of GLUES show that there is an emergent emancipation of population development from the environment with increasing population and innovation \citep{Lemmen2010rgzm,Lemmen2012,Lemmen2011}.
 
GLUES mathematically resolves the dynamics of population density and three population-averaged characteristic sociocultural traits:  technology $T_{A}$, share of agropastoral activities $C$, and economic diversity $T_{B}$.  These are defined for preindustrial societies as follows:

\begin{enumerate} 
\item Technology $T_{A}$ is a trait which describes the efficiency of food procurement---related to both foraging and farming---and improvements in health care.  In particular, technology as a model describes the availability of tools, weapons, and transport or storage facilities. It aggregates over various relevant characteristics of early societies and also represents social aspects related to work organization and knowledge management. It quantifies improved efficiency of subsistence, which is often connected to social and technological modifications  that run in parallel. An example is the technical and societal skill of writing as a means for cultural storage and administration, with the latter acting as a organizational lubricant for food procurement and its optimal allocation in space and among social groups. $T_{A}$ is labour dependent.
\item  A second model variable $C$ represents the share of farming and herding activities, encompassing both animal husbandry and plant cultivation. It describes the allocation of energy, time, or manpower to agropastoralism with respect to the total food sector. 
\item Economic diversity $T_{B}$ resolves the number of different agropastoral economies available to a regional population.  This trait is in the full model closely tied to regional vegetation resources and climate constraints; in this reduced model, it denotes a labour-independent technology.  A larger economic diversity offering different niches for agricultural or pastoral practices enhances the reliability of subsistence and the efficacy in exploiting heterogeneous landscapes.
\end{enumerate}

The temporal change of each of these characteristic traits follows the direction of increased benefit for success (i.e.\ growth) of its associated population (Appendix \refeq{eq:gad}); this concept had been derived for genetic traits in the works of \citet{Fisher1930}, and was recently more stringently formulated by Metz and colleagues \citep{Metz1992,Kisdi2010} as adaptive dynamics (AD).  In AD, the population averaged value of a trait changes at a rate which is proportional to the gradient of the fitness function evaluated at the mean trait value.  The AD approach was extended to functional traits of ecological communities \citep{Wirtz1996eve,Merico2009}, and was first applied to cultural traits of human communities by \citet{Wirtz2003gdm}.

The adaptive coevolution of the food production system $\{T_{A},T_{B},C\}$  and population $P$ (Appendix \refeqs{eq:dp}{eq:si}), which is at the heart of this model's implementation, had also been found empirically by \citet[][p. 15]{Boserup1981}: ``The close relationship which exists today between population density and food production system is the result of two long-existing processes of adaptation.
On the one hand, population density has adapted to the natural conditions for food production [];
on the other hand, food supply systems have adapted to changes in population density.''

\section{Innovation in the transition to agriculture}

\begin{figure}
\centering
\includegraphics[width=\ifreview1\else1\fi\hsize]{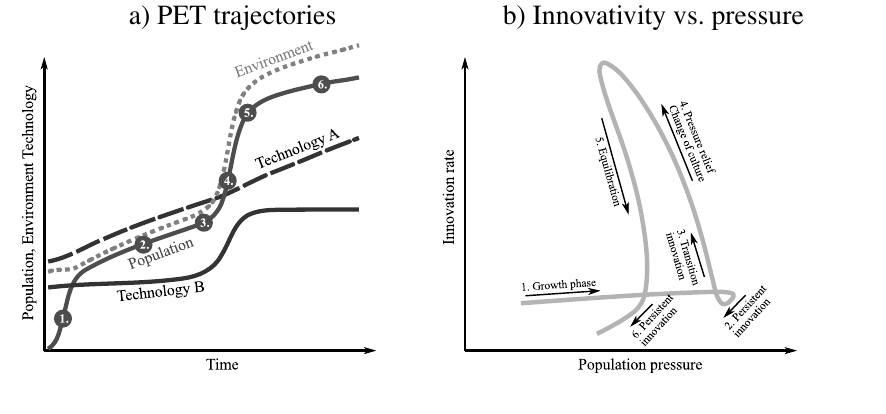}
\caption{Trajectories of population $P$, environment $E$, and technologies $T_{A},T_{B}$ (panel a) and phase diagram of innovation rate versus population pressure (panel b) from a simulation with a simplified version of the Global Land Use and technological Evolution Simulator.  The trajectories describe the temporal evolution of population density, capacity denoted as environment, a labour dependent technology $T_{A}$, and a labour-independent technology $T_{B}$.  Numbers identify the different stages of development in the both diagrams.  In the phase diagram b), the innovation rate, derived as the cumulative change in $T_{A}+T_{B}$, is shown in relation to population pressure, calculated as $1-E+P$.}
\label{fig:glues}
\end{figure}

The outcome of the coevolutionary model simulation with the reduced GLUES is shown in \reffig{fig:glues}.  I  divided both the trajectories (temporal evolution of state variables, panel~a) and the the phase space (panel~b) into six stages, which I discuss below.
\begin{enumerate}
\item Growth phase: Starting from a Malthusian perspective, and looking only at population and environment (quantified here as the ecosystem capacity, i.e. the ratio of birth over mortality terms in the growth rate \refeq{eq:r}), population grows towards its capacity with diminishing returns as $P$ approaches $E$;  this first phase spans only a short period of time but covers a large area in phase space;
\item Persistent innovation in technology $T_A$ and associated investments in tool making and administration allow sustained slow growth of population $P$ and alleviates the built-up population pressure;  in contrast to the growth phase, the phase space coverage is very small while the temporal extent of this phase is large;
\item Transition phase: rapid innovation in a labour-independent technology $T_{B}$ (e.g.~domestication successes) leads to
\item Pressure relief, but induces also a change in culture (not shown); 
\item Equilibration: Innovation slows but has led to a wider gap between $P$ and $E$ because of the investments made in manufacturing and organization during the transition: accordingly, population pressure increases more slowly and up to a lower value than in the growth phase (1.).  
\item Persistent innovation: corresponds to phase (2.) and is again characterized by persistent innovation in technology $T_{A}$ and a slow population pressure relief.
\end{enumerate}

What can be learned about the relationship between population pressure and innovativity from \reffig{fig:glues}? (i)~Innovation is greatest at high population pressure. (ii)~In this model there is always innovation, at no time is technology change negative. (iii)~The relationship between innovativity and population pressure changes profoundly during the foraging-farming transition; three different regimes can be identified: (i)~a positive relationship where acceleration of innovation corresponds to population pressure increases (phases 1.,~2.,~6.), (ii)~a negative relationship with pressure relief during accelerating innovation (phase~3.,~4.), and (iii)~a negative relationship with deceleration of innovativity at increasing pressure (phase 5.).

% Todo elaborate in more detail this is the main conclusion
A superficial analysis would find that population pressure is the motor of innovation in this example: population increase seemingly precedes the stepwise technological change (\reffig{fig:glues}a). Only a detailed look at the phase space (Figure 2b)---especially at the transition phases~2. and~3.---shows that innovativity decelerates at very high population pressure and that the largest innovation occurs slightly below the highest population pressure. In fact, the driver in the transition depicted here is not population, but technology\footnote{There would be no evolution of $T$ without $P$ due to the coevolutionary definition of the system.  The dynamics of $T$, however, leads the dynamics of $P$ at the foraging farming transition.}. Only the different coevolutionary time scales of population growth (fast) and innovation (slow) yield the seemingly Boserupian, i.e., population driven, response.

The same mathematical model---plus spatial and biogeographic aspects---has been used to successfully simulate the many transitions to agriculture in Neolithic Europe \citep{Lemmen2011}, with good agreement with the radiocarbon record. Also there, the transitions appear Boserupian with critical innovations occurring at high population pressure. If the numerical analysis had not been available (and proved that this is in fact technology driven), such as it is in the discretely sampled data from observations of technological change, one would have to have come to the erroneous conclusion that this type of innovation was population driven.

\section{Conclusion}

I presented a reduced version of the Global Land Use and technological Evolution Simulator, a numerical model which is capable of realistically simulating regional foraging-farming transitions worldwide. The simulated---and possibly also observed---transitions are seemingly Boserupian, i.e., population driven: innovation is greatest when population pressure is high. Analytical examination of the model, however, shows that technological change is the driver, and that in the context of a simplified version of  \citeauthor{Boserup1996}'s (1996) framework in development theory the model should be classified as Neomalthusian. I thus demonstrated that Boserupian appearance may be based on Malthusian assumptions; I caution not to infer too quickly a Boserupian mechanism for an observed real world system when its dynamics appears to be population pressure driven.

\appendix
\section*{Appendix: the reduced GLUES model}

A coevolutionary system of population $P$ and characteristic traits $X\in\{T_{A},T_{B},C\}$ is defined by the evolution equations
\begin{eqnarray}
\label{eq:dp}
\dd{P}{t} &=& P\cdot r\\
\dd{X}{t} &=& \delta_{X}\cdot \del{r}{X}, 
\label{eq:gad}
\end{eqnarray}
where $r$ denotes the specific growth rate of population $P$, and the $\delta_{X}$ are variability measures for each $X$.
Growth rate $r$ is defined as 
\begin{equation}
r =  \mu\cdot(1-\omega T_{A})\cdot(1-\gamma\sqrt{T_{A}} P)\cdot\si-\rho\cdot T^{-1}_{A}\cdot P,
\label{eq:r}
\end{equation}
with coefficients $\mu, \rho, \omega, \gamma$.  In this formulation, the positive term including food production SI is modulated by labour loss for administration $(-\omega T_{A})$ and by overexploitation of the environment $(-\gamma\sqrt{T_{A}} P)$.   Food production depends on the cultural system $C$ and available technologies as follows:

\begin{equation}
SI = (1-C)\cdot \sqrt(T_{A}) + C\cdot T_{A}\cdot T_{B},
\label{eq:si}
\end{equation}
where the left summand denotes foraging activities and the right summand agropastoral practice.

To produce the results for \reffig{fig:glues}, I assumed the following parameter values: 
$\mu=\rho=0.004$,
$\omega=0.04$,
$\gamma=0.12$,
$\delta_{T_{A}}=0.025$,
$\delta_{T_{B}}=0.9$;  a variable $\delta_{C}=C\cdot(1-C)$; and initial values for $P_{0}=0.01$, $T_{A,0}=1.0$, $T_{B,0}=0.8$, and $C_{0}=0.04$.

\paragraph{Acknowledgments.}
This study was partly funded by the German National Science Foundation (DFG priority project~1266 Interdynamik) and by the PACES program of the Helmholtz Gemeinschaft. The paper received great stimulus from discussions during the Ester Boserup Conference 2010---A Centennial Tribute: Long-term trajectories in population, gender relations, land use, and the environment, November 15--17, 2010 in Vienna, Austria. I received helpful comments from two anonymous reviewers. GLUES is free and open source software and can be obtained from \href{http://glues.sourceforge.net/}{http://glues.sourceforge.net/}.

\ifreview\newpage\else\fi
\bibliographystyle{agufzj}
\bibliography{mendeley,journal-macros,glues}
\end{document}